\documentclass[11pt,american,english]{article}
\usepackage[T1]{fontenc}
\usepackage{geometry}
\usepackage{amsmath}
\usepackage{babel}
\usepackage{graphics}
\usepackage{setspace}
\usepackage{amssymb}

\makeatletter

\providecommand{\LyX}{L\kern-.1667em\lower.25em\hbox{Y}\kern-.125emX\@}

\usepackage{amsfonts}
\usepackage{latexsym}

\makeatother
\begin{document}
\hfill{}UCL-IPT-01-14

\begin{center}

\vspace{1cm}

{\LARGE Gravitational dipole radiations from binary systems}

\vspace{1cm}

\def \thefootnote{\fnsymbol{footnote}}

{\large J.-M.Gérard and Y.Wiaux\footnote{FNRS Research Fellow (email:
ywiaux@fyma.ucl.ac.be)}}

\renewcommand{\thefootnote}{\arabic{footnote}}

\setcounter{footnote}{0}

\vspace{1cm}

Institut de Physique Théorique

Université catholique de Louvain

B-1348 Louvain-la-Neuve, Belgium

\vspace{1cm}

\end{center}

\begin{abstract}
We investigate the possibility of generating sizeable dipole radiations
in relativistic theories of gravity. Optimal parameters to observe
their effects through the orbital period decay of binary star systems
are discussed. Constraints on gravitational couplings beyond general
relativity are derived.
\end{abstract}

\section{Introduction}

General relativity predicts dominant quadrupole gravitational radiations.
However, alternatives to Einstein's theory actually extend the concept
of gravitational charge beyond the mass, leading, in principle, to
dipole radiations.

In this paper, we wish to investigate the possibility of generating
sizeable dipole radiations, in the framework of two minimal, but conceptually
important, extensions of general relativity. First, we emphasize that
a violation of the equivalence principle provides us with a rather
powerful indicator for gravitational dipole radiations. In section
\ref{sectionBD}, we contemplate the dominance of dipole radiations
in the Brans-Dicke scalar-tensor theory where the additional gravitational
charge is related to the gravitational binding energy of compact bodies.
Section \ref{sectionKaluza} is devoted to a Kaluza-inspired vector-tensor
theory where the new gravitational charge is proportional to the extra
dimensional velocity. In the last section, we consider binary pulsars
as indirect searching tools for gravitational dipole radiations. Monitoring
the orbital period decrease rate of binary systems leads to limits
on either the scalar or vector charges involved. In particular, we
advocate realistic tight constraints through the analysis of existing
neutron star - white dwarf binary pulsars. The resulting bounds on
a scalar gravitational coupling might actually compete with the expected
level of precision of future satellite experiments dedicated to probing
this coupling.

\section{Gravitational charges and the equivalence principle \label{sectiongravicharge}}

In electromagnetism, the classical Larmor result \cite{Jac-TB62,Lan-TB66}
for the dipole radiation rate due to cyclic motions is given by\begin{equation}
\label{int-dipem}
-\frac{dE}{dt}=\frac{1}{6\pi c^{3}}\ddot{D}^{i}\ddot{D}^{i}\, ,
\end{equation}
with \( D^{i}=\sum _{a}q_{a}r^{i}_{a} \), the dipole moment of the
source and \( q \), the electric charge. In Einstein's tensor theory,
the analogous gravitational charge is the mass \( m \) itself. Consequently,
the conservation of momentum in the non-relativistic limit implies
that dipole radiations are forbidden in general relativity and the
\( v^{2}/c^{2} \)-suppressed quadrupole radiations dominate, \( v \)
being the average velocity in the (binary) system under scrutiny.
The possibility to generate dipole radiations in relativistic theories
of gravity beyond general relativity is therefore challenging from
a phenomenological point of view.

\begin{figure}[hbtp]
{\centering \resizebox*{!}{5cm}{\includegraphics{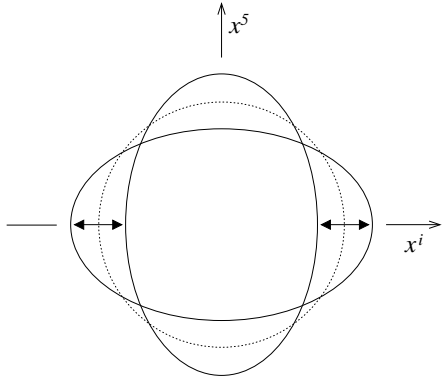}} \par}

\caption{\label{figure}Dipole as projection of extra dimensional quadrupole. }
\end{figure}

An intuitive way to ensure such a dipole radiation is to immerse Einstein's
theory of gravity in a five dimensional spacetime%
\footnote{Our convention for the metric signature is as follows: \( (+,-,-,-...) \).
}: a projection of the quadrupole motion on our \( 4 \)-dimensional
spacetime corresponds then to a dipole one (see Fig.\ref{figure}).
\\
Indeed, the action for the geodesic motion of a test particle in five
dimensions,\begin{equation}
\label{int-1}
S=-c\int \, m\, ds^{(5)}\, ,
\end{equation}
reduces to \begin{equation}
\label{int-actionvec}
S\simeq -\int \, \{mc+(m\frac{u^{5}}{c})\, u^{\mu }g_{5\mu }\}\, ds
\end{equation}
in the weak field approximation, if \( ds=cd\tau  \) is the \( 4 \)-dimensional
infinitesimal length (\( u^{\mu }=dx^{\mu }/d\tau  \) and \( u^{5}=dx^{5}/d\tau  \))
and for \( u^{5}<c \). The second term in this action corresponds
to a vector gravitational interaction similar to electromagnetism.
The appearance of a second gravitational charge \( (mu^{5}/c) \)
associated with the \( 4 \)-dimensional vector field \( g_{5\mu }(\vec{x},t) \)
circumvents the momentum conservation law. It allows for dipole radiations
if particles propagate in extra dimensions with constant intrinsic
velocities. Such a possibility will be considered in details in section
\ref{sectionKaluza}. For the time being, we simply notice that the
reduced action implies the following equations of motion\begin{equation}
\label{int-2}
m\dot{u}^{\alpha }+m\Gamma ^{\alpha }_{\mu \nu }u^{\mu }u^{\nu }\simeq c^{2}(m\frac{u^{5}}{c})\, (g^{\, \rho ,\alpha }_{5}-g^{\, \alpha ,\rho }_{5})\frac{u_{\rho }}{c}\, .
\end{equation}
Consequently, the acceleration of a body in a freely falling frame
is non zero since we obtain \begin{equation}
\label{int-3}
\vec{a}-\vec{g}\simeq -c^{2}(\frac{v^{5}}{c})\vec{\nabla }g_{50}\, ,
\end{equation}
in the non-relativistic limit. The right-hand side term being of gravitational
nature, we face here a violation of the equivalence principle, i.e.,
a violation of the universality of free fall. We therefore have established
a connection between the existence of dipole radiations and the violation
of the equivalence principle, thus generalizing a conjecture made
in \cite{Wil-TB93}. 

Let us take advantage of this link to single out another class of
relevant metric theories, in which the mass of particles would explicitly
depend on an additional scalar gravitational field \( \phi  \): \begin{equation}
\label{int-actionscal}
S=-c\int \, m(\phi )\, ds\, .
\end{equation}
In the weak field approximation, this action for matter particles
becomes\begin{equation}
\label{int-actionscallin}
S\simeq -c\int \, \left\{ m+(ms)\, \phi \right\} \, ds\, ,
\end{equation}
with \( s=(d\ln m/d\phi )_{|0} \), and the inferred equations of
motion are\begin{equation}
\label{int-eqscallin}
m\dot{u}^{\alpha }+m\Gamma ^{\alpha }_{\mu \nu }u^{\mu }u^{\nu }\simeq c^{2}(ms)\, \phi _{,\mu }(g^{\mu \alpha }-u^{\mu }u^{\alpha }/c^{2})\, .
\end{equation}
In the non-relativistic limit, \begin{equation}
\label{int-eqscallinclas}
\vec{a}-\vec{g}\simeq -c^{2}s\vec{\nabla }\phi \, ,
\end{equation}
and, once again, the existence of the second gravitational charge
\( (ms) \) implies both a violation of the equivalence principle
(see Eq.(\ref{int-eqscallinclas})) and dipole radiations (see Eq.(\ref{int-actionscallin})).

Additional scalar and vector gravitational fields seem to be unavoidable
in quantum unifications of the four known fundamental interactions.
In the next two sections, we will focus on popular prototypes of scalar-tensor
and vector-tensor models of gravity resulting from String Theory and
Supergravity, respectively.

\section{Brans-Dicke theories \label{sectionBD}}

The Brans-Dicke theory \cite{Bra-Dic} is the minimal extension of
general relativity including a second gravitational field. It is however
already sophisticated enough, as we indicated in the previous section,
to encounter an equivalence principle violation and to allow for large
dipole radiations.

\subsection{Scalar-tensor theory for compact bodies}

In the context of scalar-tensor theories, the mass of extended bodies
will naturally depend on the scalar field's background value. Indeed,
considering a freely falling frame (hence switching off the metric
field), the scalar interaction survives. The internal structure of
a body depends on it, and therefore its overall mass too. As proposed
by Eardley \cite{Ear75}, we will treat a compact (i.e. self-gravitating)
body in this framework as being poinlike, though with an explicit
scalar dependence in its mass. The corresponding action (including
now dynamics for the gravitational fields) of Brans-Dicke theory reads\begin{equation}
\label{bd-1}
S^{bd}=-\frac{c^{3}}{16\pi }\int \sqrt{g}d^{4}x\, \left( \Phi R-\frac{\omega }{\Phi }\Phi ^{,\alpha }\Phi _{,\alpha }\right) -c\int \, m(\Phi )\, ds\, ,
\end{equation}
where \( g \) is the absolute value of the determinant of the metric
\( g_{\mu \nu } \), \( \Phi  \) is the massless scalar field and
\( \omega  \), a dimensionless parameter. We are however only interested
here in the linearized version of this action. The constant \( \Phi _{0} \)
standing for the cosmological background value of the scalar field,
let us define the metric and scalar perturbations as follows: \foreignlanguage{american}{}\begin{eqnarray}
g_{\mu \nu } & = & \eta _{\mu \nu }+\sqrt{2c\kappa '}(h_{\mu \nu }-\alpha \eta _{\mu \nu }\varphi )\nonumber \label{bd-2} \\
\frac{\Phi }{\Phi _{0}} & = & 1+\alpha \sqrt{2c\kappa '}\varphi \, .\label{bd-3} 
\end{eqnarray}
The decoupling and canonical normalization%
\footnote{Notice that the somewhat unusual additional \( \sqrt{c} \) factor
in the normalization of the gravitational fields allows us to transfer
the whole dependence on the light velocity into the couplings. The
\( 1/c \) counting will then be obvious when evaluating the intensity
of radiation.
} of the corresponding kinetic terms in the resulting theory, \begin{eqnarray}
S^{grav}_{(bd)} & = & \int d^{4}x\, \left[ \left( \frac{1}{4}\partial _{\mu }h^{\alpha \beta }\partial ^{\mu }h_{\alpha \beta }-\frac{1}{4}\partial _{\mu }h\partial ^{\mu }h+\frac{1}{2}\partial ^{\mu }h\partial ^{\nu }h_{\mu \nu }-\frac{1}{2}\partial _{\nu }h^{\mu \nu }\partial ^{\rho }h_{\mu \rho }\right) \right. \nonumber \\
 &  & \left. +\left( \frac{1}{2}\partial _{\mu }\varphi \partial ^{\mu }\varphi \right) \right] \, ,\label{bd-Sgrav} 
\end{eqnarray}
are implemented for \( \kappa '\equiv 8\pi G'/c^{4} \), (\( G'\equiv 1/\Phi _{0} \)
being the bare tensor coupling constant) and \begin{equation}
\label{bd-alphaomega}
\alpha ^{2}=\frac{1}{2\omega +3}\, .
\end{equation}
With such a normalization, the tensor and scalar propagators are given
by\begin{equation}
\label{bd-propten}
G^{\mu \nu ,\alpha \beta }(q)=\frac{\eta ^{\mu \alpha }\eta ^{\nu \beta }+\eta ^{\mu \beta }\eta ^{\nu \alpha }-\eta ^{\mu \nu }\eta ^{\alpha \beta }}{q^{2}+i\varepsilon }
\end{equation}
and \begin{equation}
\label{bd-propsca}
\Delta _{F}(q)=\frac{1}{q^{2}+i\varepsilon }\, ,
\end{equation}
respectively, for a transfer momentum \( q \). Notice that the tensor
propagator is the same as in general relativity.

The linearized action for matter (assuming non-relativistic motion)
gives the following interaction terms: \begin{equation}
\label{bd-Sint}
S^{int}_{(bd)}=-\sqrt{\frac{\kappa '}{2c}}\int d^{4}x\, \left[ h^{\mu \nu }T_{\mu \nu }-\alpha \varphi T\left( 1-2s\right) \right] \, ,
\end{equation}
where \( T^{\mu \nu }=\rho \, v^{\mu }v^{\nu } \) \( (\rho \equiv m\, \delta ^{3}(\vec{x}-\vec{x}(t))) \)
is the non-relativistic expression of the energy-momentum tensor for
ordinary matter and \( T=\rho c^{2} \), its trace. The constant \( \alpha  \)
is identified with the scalar coupling to matter and, consequently,
we recover general relativity in the limit \( \omega \rightarrow \infty  \)
(see Eq.(\ref{bd-alphaomega})). The new gravitational charge \( s \)
is the sensitivity of the variable mass with respect to the scalar
field:\begin{equation}
\label{bd-6}
s\equiv \frac{\partial \ln m}{\partial \ln \Phi }_{|\Phi _{0}}\, .
\end{equation}

\subsection{Strong equivalence principle violation}

The effective gravitational coupling \( G_{12} \) between two compact
bodies (labeled by \( 1 \) and \( 2 \)) here follows from the virtual
exchange of a tensor \( (T) \) and a scalar \( (S) \) gravitational
fields. In the non-relativistic limit, the interaction potential is
given by the scattering amplitude \( \mathcal{M}\equiv \mathcal{M}^{T}+\mathcal{M}^{S} \)
of the two particles. In the limit of static bodies \( (T^{\mu \nu }\equiv \rho c^{2}\delta _{\, 0}^{\mu }\delta _{\, 0}^{\nu }) \),
we get from Eqs.(\ref{bd-propten}-\ref{bd-Sint}),\begin{equation}
\label{bd-5}
\mathcal{M}=\mathcal{N}c^{4}\, \left[ 1+\alpha ^{2}(1-2s_{1})(1-2s_{2})\right] G'\frac{m_{1}m_{2}}{-\vec{q}^{2}}\, ,
\end{equation}
 where \( \mathcal{N} \) is a standard normalization factor, and
\( \vec{q} \), the spatial component of the transfer momentum. The
corresponding interaction potential is therefore the Newton potential
with\begin{equation}
\label{bd-G12}
G_{12}=\left[ 1+\alpha ^{2}(1-2s_{1})(1-2s_{2})\right] G'\, .
\end{equation}
In the case of test particles, this effective coupling constant reduces
to \begin{equation}
\label{bd-G}
G=(1+\alpha ^{2})G'\, ,
\end{equation}
 which defines the universal Newton coupling constant \( G \) in
terms of the bare coupling \( G' \). This relation allows us to interpret
the charge \( s \) as the ratio of the internal gravitational binding
energy \( \Omega _{gr} \) of the body to its mass \( m \):\begin{eqnarray}
s & = & -\frac{\partial \ln m}{\partial \ln G}\nonumber \label{bd-7} \\
 & = & \frac{|\Omega _{gr}|}{mc^{2}}\, .\label{bd-8} 
\end{eqnarray}

Before tackling the issue of dipole radiations, we would like to emphasize
that, here we have in fact a violation of the strong equivalence principle,
i.e., the violation of the universal geodesic motion for self-gravitating
bodies only. Indeed, the free fall of a compact body towards a non
compact source \( (s_{1}=0) \) is slowed down proportionally to its
own sensitivity \( (s_{2}) \): from Eqs.(\ref{bd-G12}) and (\ref{bd-G}),
we obtain\begin{equation}
\label{bd-9}
G_{12}=(1-\xi s_{2})G\, ,
\end{equation}
with \begin{equation}
\label{bd-10}
\xi =\frac{2\alpha ^{2}}{(1+\alpha ^{2})}\, ,
\end{equation}
the effective scalar coupling responsible for what is called the {}``Nordtvedt
effect'' \cite{Nor68,Wil-TB93}. This effect, anomalous with respect
to general relativity, is responsible for a polarization towards the
Sun of the Moon's orbit around the Earth. The Lunar Laser Ranging
experiment, monitoring the motion of the Moon with an impressive current
precision of the order of one centimeter, puts a tight bound on the
amplitude of such a polarization. Working in the restrictive limit
of the Brans-Dicke theory, this amounts to constraining the effective
scalar coupling to extremely small values \cite{Wil00}:\begin{equation}
\label{bd-11}
\xi \leq 6\, 10^{-4}\, .
\end{equation}

\subsection{Scalar dipole}

As already noticed in the first paragraph of this section, (see Eq.(\ref{bd-Sint})),
the source of the scalar field is \( m(1-2s) \) rather than the mass
\( m \) itself. Consequently, monopole radiations are still forbidden,
but dipole ones (induced by the charge \( s \)) will appear in the
same way as in electromagnetism. Let us therefore make the identification
of the scalar Brans-Dicke interaction proportional to the sensitivity
\( s \) with the vector interaction present in the Maxwell action
\begin{equation}
\label{bd-12}
S^{em}=\int d^{4}x\, \left[ -\frac{1}{4}F_{\mu \nu }F^{\mu \nu }-\frac{1}{c^{3/2}}A_{\mu }j^{\mu }\right] \, ,
\end{equation}
where \( j^{\mu }=q\delta ^{3}(\vec{x}-\vec{x}(t))\, v^{\mu } \).
If we perform the substitution\begin{equation}
\label{bd-13}
\frac{q_{a}}{m_{a}}=\sqrt{16\pi G'}\, \alpha s_{a}
\end{equation}
in our basic Eq.(\ref{int-dipem}), taking into account a polarization
correction factor \( 1/2 \) when identifying \( A_{\mu }v^{\mu }/c \)
with \( \varphi  \), we readily obtain the scalar dipole radiation
rate\begin{equation}
\label{bd-tauxenerscal}
-\frac{dE_{S}^{dip}}{dt}=\frac{2}{3}\xi \frac{G}{c^{3}}{\ddot{D}_{(s)}}^{i}{\ddot{D}_{(s)}}^{i}\, ,
\end{equation}
 with \( D_{(s)}^{i}\equiv \int d^{3}\vec{x}'\, \rho s\, {x'}^{i} \).
This is precisely the result of the more standard procedure of developping
the radiative solution for \( \varphi  \) in multipole moments. In
the particular case of a binary source in Keplerian motion, the dipole
moment is given by \( \vec{D}_{(s)}=m_{1}s_{1}\vec{r}_{1}+m_{2}s_{2}\vec{r}_{2} \)
and we get, to first significant order in the scalar coupling \( \xi  \)
, the average rate\begin{equation}
\label{bd-tauxenerscabin}
\frac{1}{T}\int _{0}^{T}\, \left( -\frac{dE_{S}^{dip}}{dt}\right) \, dt\equiv \left\langle -\frac{dE_{S}^{dip}}{dt}\right\rangle _{K}=\frac{2}{3}\xi \frac{G^{3}}{c^{3}}\frac{\mu ^{2}M^{2}}{a^{4}}\left( s_{1}-s_{2}\right) ^{2}g(e)\, ,
\end{equation}
with \( \mu  \) and \( M \), the reduced and total masses respectively,
and \( a \), the semi major axis of the classical orbit. The enhancement
factor \( g(e)=\left( 1+e^{2}/2\right) /\left( 1-e^{2}\right) ^{5/2} \)
is a function of the orbital eccentricity \( e \).

The second kind of emission%
\footnote{We will neglect the scalar quadrupole (as well as a scalar monopole
not even encountered by our linearized theory) as it would be of order
\( O(\xi /c^{5}) \), hence negligible compared to the tensor quadrupole,
of order \( O(1/c^{5}) \).
} coming out in the context of these theories is the standard quadrupole
tensor radiation \cite{Lan-TB66,Wei-TB72,Pet-Mat63} of general relativity
(with again \( G' \) substituted for \( G \)):\begin{equation}
\label{bd-tauxenertens}
-\frac{dE_{T}^{quad}}{dt}=\frac{G'}{5c^{5}}\dddot D^{ij}\dddot D_{ij}\, ,
\end{equation}
with \( D^{ij}\equiv \int d^{3}\vec{x}'\, \rho \, ({x'}^{i}{x'}^{j}-\frac{1}{3}\delta ^{ij}r^{'2}) \).
The Keplerian motion of a two-body system leads to%
\footnote{Here \( G' \) and \( G_{12} \) are set equal to \( G \). Once more,
this corresponds to neglecting small corrections of order \( O(\xi /c^{5}) \).
} the average rate \begin{equation}
\label{bd-tauxenertensbin}
\left\langle -\frac{dE_{T}^{quad}}{dt}\right\rangle _{K}=\frac{32}{5}\frac{G^{4}}{c^{5}}\frac{\mu ^{2}M^{3}}{a^{5}}f(e)\, ,
\end{equation}
with the enhancement factor \( f(e)=\left( 1+73/24e^{2}+37/96e^{4}\right) /\left( 1-e^{2}\right) ^{7/2} \). 

As already emphasized, dipole radiations (of order \( O(1/c^{3}) \))
could a priori dominate quadrupole ones (of order \( O(1/c^{5}) \)).
But, before discussing phenomenological implications, let us turn
our attention to the other previously mentioned theoretical frame
where dipole radiations are also expected.

\section{Kaluza theories \label{sectionKaluza}}

We have considered the possibility of dipole radiations arising from
the coupling of a \emph{scalar} gravitational field to the \emph{sensitivity}
of compact systems. Could this radiation rather come from the coupling
of a \emph{vector} field of gravitation to an \emph{extra dimensional
velocity}, as already suggested in section \ref{sectiongravicharge}
? 

In 1921, Kaluza \cite{Kal21} introduced for the first time the seminal
concept of extra dimensions in a superb attempt to unify gravitation
and electromagnetism. In this theory, classical matter particles are
allowed to propagate in a \( 5 \)-dimensional spacetime and the momentum
\( (mv^{5}) \) in the fifth direction is interpreted in terms of
the electric charge \( (q) \) \begin{equation}
\label{ed-identifqv}
\frac{q}{m}=\sqrt{16\pi G}(\frac{v^{5}}{c})\, .
\end{equation}
However, this identification constrains the electric charge \( q \)
to irrationally small values for elementary particles, which led to
abandon temporarily the concept of extra dimensional spacetime. Here,
we would like to preserve the gravitational nature of the induced
vector interaction and to analyze the observable consequences of the
associated charge for macroscopic phenomena.

\subsection{Gravity in \protect\( 4+\delta \protect \) dimensions}

The Einstein theory in \( 4+\delta  \) dimensions is given by the
action\begin{equation}
\label{ed-01}
S_{(4+\delta )}=-\frac{1}{2c\kappa _{(4+\delta )}}\int d^{(4+\delta )}x\, \sqrt{\hat{g}}\hat{R}+\int d^{(4+\delta )}x\, \sqrt{\hat{g}}\mathcal{L}_{mat}\left( \Psi ,\, \hat{g}_{MN}\right) \, ,
\end{equation}
where \( \kappa _{(4+\delta )}\equiv 8\pi G_{(4+\delta )}/c^{4} \).
Here, \( G_{(4+\delta )} \) stands for the \( (4+\delta ) \)-dimensional
coupling constant and \( \hat{g}_{MN} \) is the metric of the \( (4+\delta ) \)-dimensional
spacetime%
\footnote{Throughout this article, we will stick to the following indices conventions:
capital indices run over all dimensions \( (0\leq M,N...\leq 4+\delta ) \),
greek indices over the usual spacetime dimensions \( (0\leq \mu ,\nu ...\leq 3) \),
latin \( ij \)-indices run over the usual spatial dimensions \( (1\leq i,j...\leq 3) \),
while latin \( mn \)-indices run over extra dimensions only \( (4\leq m,n...\leq 4+\delta ) \).
}. If one defines the metric perturbation as \( \hat{g}_{MN}=\eta _{MN}+\sqrt{2c\kappa _{(4+\delta )}}\hat{h}_{MN} \),
the corresponding linearized theory (which is the one of interest)
reads\begin{eqnarray}
S^{quad}_{(4+\delta )} & = & \int d^{(4+\delta )}x\, \left[ \left( \frac{1}{4}\partial _{M}\hat{h}^{AB}\partial ^{M}\hat{h}_{AB}-\frac{1}{4}\partial _{M}\hat{h}\partial ^{M}\hat{h}+\frac{1}{2}\partial ^{M}\hat{h}\partial ^{N}\hat{h}_{MN}\right. \right. \nonumber \\
 &  & \left. \left. -\frac{1}{2}\partial _{N}\hat{h}^{MN}\partial ^{R}\hat{h}_{MR}\right) -\sqrt{\frac{\kappa _{(4+\delta )}}{2c}}\hat{h}^{MN}T_{MN}\right] \, ,\label{actionlingravbulk} 
\end{eqnarray}
where \( T_{MN} \) is the stress tensor for ordinary matter. For
simplicity, we will assume that the extra dimensions are compactified.
The compactification on a torus (with identical radii \( R \)) implies
that each field will decompose into modes \( \hat{h}^{(\vec{n})}_{MN}(x^{\mu }) \).
Only modes with the same \( \vec{n} \) will couple to each other.
Consequently, the integration over the \( \delta  \) extra dimensions
in the action simply leads to a volume factor \( V^{\delta } \) which
will renormalize the gravitational constant when going from \( 4+\delta  \)
to \( 4 \) dimensions: \( G_{4}\equiv G_{4+\delta }/V^{\delta }\, . \)
The structure of the \( 4 \)-dimensional interaction (fields and
couplings) may be derived through the following {}``metric reduction''
\cite{Han-Lyk-Zha99}:\begin{equation}
\label{ed-02}
\eta _{MN}+\sqrt{2c\kappa _{(4+\delta )}}\hat{h}_{MN}=\eta _{MN}+\sqrt{2c\kappa _{4}}\left[ \begin{array}{cc}
h_{\mu \nu }-\alpha \eta _{\mu \nu }\varphi  & A_{\mu n}\\
A_{m\nu } & -2\alpha \varphi _{mn}
\end{array}\right] \, ,
\end{equation}
 where \( \kappa _{4} \) is the \( 4 \)-dimensional coupling. The
scalar \( \varphi =\delta ^{mn}\varphi _{mn} \) is the trace of the
\( \delta (\delta +1)/2 \) \( 4 \)-dimensional scalar fields, \( A_{\mu n} \)
are \( \delta  \) vector fields. The parameter \( \alpha  \) is
again introduced to normalize the kinetic term for the scalar \( \varphi  \).
This process of {}``metric reduction'' is rather long and awkward,
as one has to operate intricate field redefinitions before correctly
identifying \( \vec{n}=0 \) modes with massless mediators and \( \vec{n}\neq 0 \)
modes with infinite (Kaluza-Klein) towers of massive mediators of
the interaction. 

However, one may find out the field content (with correct couplings)
much more easily via a {}``propagator reduction''. Let us first
consider how the \( (4+\delta ) \)-interaction is mediated and decompose
the propagator in a second step. In \( 4+\delta  \) dimensions, there
is only one graviton. From Eq.(\ref{actionlingravbulk}), it is straightforward
to define its propagator as\begin{equation}
\label{ed-03}
G^{MN,AB}(q)=\frac{\eta ^{MA}\eta ^{NB}+\eta ^{MB}\eta ^{NA}-\frac{2}{2+\delta }\eta ^{MN}\eta ^{AB}}{q_{(4+\delta )}^{2}+i\varepsilon }\, ,
\end{equation}
where \( q_{(4+\delta )} \) is the momentum transfer in \( 4+\delta  \)
dimensions. In order to understand the couplings to the \( 4 \)-dimensional
matter particles, we may once again consider the scattering of two
particles by the gravitational interaction. For matter propagating
in our \( 4 \)-dimensional world, the corresponding amplitude reads
\begin{eqnarray}
\mathcal{M} & = & \mathcal{N}\, \sum _{q_{(4+\delta )}}\frac{1}{V^{\delta }}G_{(4+\delta )}\, T^{(1)}_{\mu \nu }G^{\mu \nu ,\alpha \beta }(q)T^{(2)}_{\alpha \beta }\nonumber \label{ed-04} \\
 & = & \mathcal{N}\, \sum _{q_{(4+\delta )}}G_{4}\, T^{(1)}_{\mu \nu }G^{\mu \nu ,\alpha \beta }(q)T^{(2)}_{\alpha \beta }\, ,\label{ed-05} 
\end{eqnarray}
with \( q_{4} \), a given transfer momentum. From compactification,
we know that each mode is associated with a mass: \( q_{(4+\delta )}^{2}=q_{4}^{2}-m_{n}^{2}c^{2} \),
where \( m_{n}c=\frac{|\vec{n}|}{R} \). The total scattering amplitude
splits then into one massless and massive terms, \begin{equation}
\label{ed-06}
\mathcal{M}=\mathcal{M}^{ST}_{0}+\sum _{n}\mathcal{M}^{ST}_{n}\, ,
\end{equation}
each of them corresponding to the interaction between matter fields
through the propagation of massless or massive gravitational mediators,
respectively.

Neglecting for a while the massive states, we may decompose the original
propagator into massless tensor and scalar propagators in \( 4 \)
dimensions, with fixed relative couplings:\begin{eqnarray}
\mathcal{M}^{ST}_{0} & = & \mathcal{N}\, \left( \frac{1}{q_{4}^{2}+i\varepsilon }\right) G_{4}\, T^{(1)}_{\mu \nu }\left( \eta ^{\mu \alpha }\eta ^{\nu \beta }+\eta ^{\mu \beta }\eta ^{\nu \alpha }-\frac{2}{2+\delta }\eta ^{\mu \nu }\eta ^{\alpha \beta }\right) T^{(2)}_{\alpha \beta }\nonumber \label{ed-07} \\
 & = & \mathcal{N}\, \left( G_{4}\, T^{(1)}_{\mu \nu }\frac{\eta ^{\mu \alpha }\eta ^{\nu \beta }+\eta ^{\mu \beta }\eta ^{\nu \alpha }-\eta ^{\mu \nu }\eta ^{\alpha \beta }}{q_{4}^{2}+i\varepsilon }T^{(2)}_{\alpha \beta }\right. \nonumber \\
 &  & \qquad \qquad \qquad \qquad \qquad \qquad +\left. \left[ \frac{\delta }{\delta +2}G_{4}\right] \, T^{(1)}\frac{1}{q_{4}^{2}+i\varepsilon }T^{(2)}\right) \, .\label{ed-diff4d} 
\end{eqnarray}
The corresponding effective action for gravitation in \( 4 \) dimensions
is given by\begin{equation}
\label{ed-09}
S^{ST}_{0}=S^{T}_{0}+S^{S}_{0}\, ,
\end{equation}
where the tensor part \( S^{T}_{0} \) is \begin{eqnarray}
S^{T}_{0} & = & \int d^{4}x\, \left[ \left( \frac{1}{4}\partial _{\mu }h^{\alpha \beta }\partial ^{\mu }h_{\alpha \beta }-\frac{1}{4}\partial _{\mu }h\partial ^{\mu }h+\frac{1}{2}\partial ^{\mu }h\partial ^{\nu }h_{\mu \nu }-\frac{1}{2}\partial _{\nu }h^{\mu \nu }\partial ^{\rho }h_{\mu \rho }\right) \right. \nonumber \\
 &  & \left. -\sqrt{\frac{\kappa _{4}}{2c}}h_{\mu \nu }T^{\mu \nu }\right] \, ,\label{ed-10} 
\end{eqnarray}
and the scalar one \( S^{S}_{0} \) reads\begin{equation}
\label{ed-11}
S^{S}_{0}=\int d^{4}x\, \left[ \left( \frac{1}{2}\partial _{\mu }\varphi \partial ^{\mu }\varphi \right) +\sqrt{\frac{\kappa _{4}}{2c}}\sqrt{\frac{\delta }{\delta +2}}\varphi T\right] \, .
\end{equation}
These are precisely the results obtained through the usual {}``metric
reduction''. They correspond to the linearized version of a Brans-Dicke
scalar-tensor theory with strong scalar coupling, \begin{equation}
\label{ed-12}
\alpha ^{2}=\frac{\delta }{\delta +2}\geq \frac{1}{3}\, ,
\end{equation}
if \( G_{4}=G' \) (see Eq.(\ref{bd-Sint}) in the limit of test bodies
\( (s=0) \)).

The massive modes%
\footnote{For these modes, the {}``propagator reduction'' method leads to:
\begin{eqnarray*}
\sum _{\vec{n}}\mathcal{M}_{\vec{n}} & = & \mathcal{N}\, \left( \sum _{\vec{n}}\frac{1}{q_{4}^{2}-m_{n}^{2}c^{2}+i\varepsilon }\right) G_{4}\, T^{(1)}_{\mu \nu }\left( \eta ^{\mu \alpha }\eta ^{\nu \beta }+\eta ^{\mu \beta }\eta ^{\nu \alpha }-\frac{2}{2+\delta }\eta ^{\mu \nu }\eta ^{\alpha \beta }\right) T^{(2)}_{\alpha \beta }\\
 & = & \mathcal{N}\, \left[ \sum _{\vec{n}}G_{4}\, T^{(1)}_{\mu \nu }\frac{\eta ^{\mu \alpha }\eta ^{\nu \beta }+\eta ^{\mu \beta }\eta ^{\nu \alpha }-\frac{2}{3}\eta ^{\mu \nu }\eta ^{\alpha \beta }}{q_{4}^{2}-m^{2}_{n}c^{2}+i\varepsilon }T^{(2)}_{\alpha \beta }+\sum _{\vec{n}}\left[ \beta ^{2}G_{4}\right] \, T^{(1)}\frac{1}{q_{4}^{2}-m_{n}^{2}c^{2}+i\varepsilon }T^{(2)}\right] \, ,
\end{eqnarray*}
with \( \beta ^{2}=(2/3)\left[ (\delta -1)/(\delta +2)\right]  \).
The corresponding \( 4 \)-dimensional theory \cite{Han-Lyk-Zha99,Giu-Rat-Wel99}
encounters one massive graviton (whose tensor structure is known to
be slightly different from the massless case given in Eq.(\ref{bd-propten}))
and one massive scalar for each mode \( \vec{n} \).
} are supposed to be heavy enough to ensure that the corresponding
Yukawa contributions do not interfere with the experimentally verified
\( 1/r \) law, above one millimiter. However, the resulting scalar-tensor
theory is totally excluded by solar system observations. In fact,
the well-known deflection of star light by the Sun is sufficient to
rule out a strong scalar coupling. Indeed, the light bending by a
heavy body is given by the scattering amplitude of a photon and a
massive scalar, through the exchange of gravitational fields. The
tracelessness of the photonic energy-momentum tensor \( (T^{(\gamma )}=0) \)
prevents from any coupling to the massless field \( \varphi  \) (see
Eq.(\ref{ed-diff4d})). Consequently, the predicted deflection angle
\( \Theta _{ST} \) is equal to the angle \( \Theta _{GR} \) of general
relativity, up to a sizeable rescaling of the effective coupling constant
(see Eq.(\ref{bd-G})): \begin{equation}
\label{ed-13}
\Theta _{ST}=\frac{G'}{G}\, \Theta _{GR}=\frac{1}{1+\alpha ^{2}}\, \Theta _{GR}\leq \frac{3}{4}\, \Theta _{GR}\, .
\end{equation}
So, we have to invoke some mechanism by which the scalar \( \varphi  \)
also acquires a mass and decouples \cite{Ark-Dim-Mar98}. This amounts
to sending the scalar coupling \( \alpha  \) to zero, providing therefore
an obviously viable model. 

Such models with {}``gravity propagating in a \( (4+\delta ) \)-dimensional
bulk and ordinary matter confined on our \( 4 \)-dimensional brane
world'' have recently been introduced to solve the hierarchy problem
of elementary particle physics \cite{Ark-Dim-Dvaorig98,Han-Lyk-Zha99,Giu-Rat-Wel99}.
In the original models, the \( \delta  \) extra dimensions are compactified,
as we have chosen to do it here. Experimental constraints still allow
these dimensions to be large, as long as they are smaller than one
millimiter (see \cite{Ark-Dim-Dvaorig98} and references therein).

Back to our concern, looking for potential sources of dipole radiations,
we abandon the so-called Large Extra Dimensional framework and allow
for matter particles themselves to propagate also in the extra dimensions.
Treating the extra dimensional velocity as a new gravitational charge
(see Eq.(\ref{int-actionvec})), we wonder whether its existence may
lead to observable consequences in terms of dipole radiations.

\subsection{Equivalence principle violation}

For simplicity, we confine the motion of bodies to one extra direction,
say \( x^{5} \). The unique vector interaction \( (V) \) easily
comes out from the {}``propagator reduction'' when considering once
more the dominant scattering amplitude for two particles (\( 1 \)
and \( 2 \)). Thanks to the extra dimensional coupling, this massless
amplitude now reads\begin{equation}
\label{ed-14}
\mathcal{M}_{0}=\mathcal{M}^{T}_{0}+\mathcal{M}^{V}_{0}
\end{equation}
where \( \mathcal{M}^{T}_{0} \) is the tensor part of \( \mathcal{M}^{ST}_{0} \)
given in Eq.(\ref{ed-diff4d}), and \begin{eqnarray}
\mathcal{M}^{V}_{0} & = & \mathcal{N}\, 4G\, T^{(1)}_{\mu i}\, \frac{\eta ^{\mu \alpha }\eta ^{ij}+\eta ^{\mu j}\eta ^{\alpha i}-\frac{2}{2+\delta }\eta ^{\mu i}\eta ^{\alpha j}}{q_{4}^{2}+i\varepsilon }\, T^{(2)}_{\alpha j}\nonumber \label{ed-15} \\
 & = & \mathcal{N}c^{2}\, 4G\, j_{(v)\mu }^{(1)}\, \frac{-\eta ^{\mu \alpha }}{q_{4}^{2}+i\varepsilon }\, j^{(2)}_{(v)\alpha }\, ,\label{ed-diff5d} 
\end{eqnarray}
where \( j_{(v)}^{\mu }=(\rho v^{5}/c)\, v^{\mu } \) defines the
vector current%
\footnote{The conservation of this current readily follows from the linearized
conservation equations in \( 4+\delta  \) dimensions \( (T^{MN}_{\hspace {4mm},N}=0) \),
after integration over the whole compactified extra dimensional volume.
} . In the non-relativistic limit, and considering static sources \( (T^{\mu \nu }\equiv \rho c^{2}\delta _{\, 0}^{\mu }\delta _{\, 0}^{\nu }) \),
the scattering amplitude is then given by\begin{equation}
\label{ed-19}
\mathcal{M}_{0}=\mathcal{N}c^{4}\, G(1-4\frac{v^{5}_{1}v^{5}_{2}}{c^{2}})\frac{m_{1}m_{2}}{-\vec{q}^{2}}\, .
\end{equation}
It corresponds to a Newtonian interaction, with a renormalized gravitational
constant \begin{equation}
\label{ed-Gvectens}
G_{12}\equiv G(1-4\frac{v^{5}_{1}v^{5}_{2}}{c^{2}})\, .
\end{equation}
This is the mathematical expression of the equivalence principle violation
in extra dimensional (vector-tensor) theories. Notice here some fundamental
differences with the case of the Brans-Dicke (scalar-tensor) theory.
On one hand, we have no clue of how the extra dimensional charge \( (v^{5}/c) \)
may a priori be linked to the nature of the body considered. In particular,
there is no reason whatsoever to focus on compact bodies only. E\"otv\"os-type
experiments should also be considered in this context. They could
tightly constrain the extra dimensional (vector) charge \cite{Su-Hec-Ade94}
of laboratory-size objects. On the other hand, just as in the case
of the strong equivalence principle violation in the Brans-Dicke theory,
an anomalous acceleration is also expected for planetary-size bodies,
though with two major conceptual differences. First, the two bodies
(\( 1 \) and \( 2 \)) considered in this case need now to be charged
(see Eq.(\ref{ed-Gvectens})). And secondly, the additional interaction
being of vector nature, this anomalous motion may either be slowed
down or accelerated. On the contrary, the analytical expressions for
scalar and vector dipole radiation rates turn out to be very similar.

\subsection{Vector dipole }

The effective vector-tensor action corresponding to the propagator
reduction given in Eq.(\ref{ed-diff5d}) reads\begin{equation}
\label{ed-17}
S^{VT}_{0}=S^{T}_{0}+S^{V}_{0}\, ,
\end{equation}
with\begin{equation}
\label{ed-18}
S^{V}_{0}=\int d^{4}x\, \left[ -\left( \frac{1}{4}F_{\mu \nu }F^{\mu \nu }\right) -\frac{1}{c^{3/2}}\sqrt{16\pi G}A_{\mu }j_{(v)}^{\mu }\right] \, .
\end{equation}
Consequently, the dipole formula readily comes out from the identification
of the extra dimensional velocity with the electromagnetic charge
(see Eq.(\ref{ed-identifqv})) in our basic Eq.(\ref{int-dipem}).
The vector dipole radiation rate therefore reads\begin{equation}
\label{ed-tauxenervect}
-\frac{dE^{dip}_{V}}{dt}=\frac{8}{3}\frac{G}{c^{3}}\ddot{D}_{(v)}^{i}\ddot{D}_{(v)}^{i}\, ,
\end{equation}
where the dipole moment is now \( D^{i}\equiv \int d^{3}x\, (\rho v^{5}/c)\, x^{i} \).
To first significant order in the velocities \( (G_{12}\rightarrow G) \),
the vector dipole radiations due to a Keplerian motion of a binary
system is therefore given by \begin{equation}
\label{ed-tauxenervectbin}
\left\langle -\frac{dE^{dip}_{V}}{dt}\right\rangle _{K}=\frac{8}{3}\frac{G^{3}}{c^{3}}\frac{\mu ^{2}M^{2}}{a^{4}}\left( \frac{v^{5}_{1}}{c}-\frac{v^{5}_{2}}{c}\right) ^{2}g(e)\, ,
\end{equation}
where \( g(e) \) is the enhancement factor already defined for the
scalar dipole (see Eq.(\ref{bd-tauxenerscabin})).

\section{Constraints from binary pulsars \label{sectioncontraintes}}

Binary pulsars are ideal laboratories for testing relativistic gravity,
thanks to the extreme precision of their radio pulse. The most famous
among them is the double neutron star system \( PSR\, B1913+16 \),
discovered by Hulse and Taylor in 1974. The analysis of radio pulse
arrival times provides us with a measurement of the decrease rate
of the system's orbital period, in agreement with the general relativity
prediction for gravitational radiations to within \( 0.4\% \) \cite{Tay93}.
Until now, the \( PSR\, B1534+12 \) is the second and last example
of a double neutron star system for which the measurement of the orbital
period decrease rate was found \cite{Sta-Tay98} to be in agreement
with general relativity, though with a less impressive precision%
\footnote{More recently, a new test of relativistic gravity was achieved \cite{Str-Bai01}
in the closest and brightest of the known binary pulsars in our galaxy,
the neutron star - white dwarf \( PSR\, J0437-4715 \). However, this
test is not provided by the analysis of radiation effects, but through
the measurement of the (range and shape of) time delay undergone by
radio pulses periodically passing close to the companion star. A similar
measurement has also been achieved for \( PSR\, B1534+12 \) leading
to test the theory to within \( 1\% \) \cite{Sta-Tay98}, that is,
a far better precision than what orbital decay analysis provides.
} of about \( 15\% \). However, in the last few years, there have
been numerous discoveries \cite{Kas00,Edw-Bai00,Cam-Kas00} of neutron
star - white dwarf systems. Notice that the birth rate of these binaries
is, according to the most recent models, at least as high as for double
neutron star binaries. Prospects for relativistic parameter measurements
in these systems, including orbital decay, are very encouraging, particularly
in the case of \( PSR\, J1141-6545 \) considered here below.

Before coming up with more phenomenological considerations, we wish
here to put forward the optimal conditions on the system's parameters
under which gravitational dipole radiations dominate. The scalar (or
vector) to tensor ratio of energy losses turns out to be simply expressed
in terms of the model-dependent gravitational charges, and of the
observable eccentricity and periastron shift (the latter including
in fact the standard \( v^{2}/c^{2} \)-quadrupole suppression factor).
From an astrophysical point of view, it is more convenient to introduce
the corresponding ratio of the orbital period decrease rates induced
by gravitational radiations. In the Keplerian approximation, these
ratios are identical. For Brans-Dicke theories, Eqs.(\ref{bd-tauxenertensbin})
and (\ref{bd-tauxenerscabin}), lead, to first order in \( \xi  \),
to \begin{equation}
\label{co-ratios}
\left\langle \frac{\dot{T}^{ST}}{\dot{T}^{RG}}\right\rangle _{K}=\left\langle \frac{\dot{E}^{ST}}{\dot{E}^{RG}}\right\rangle _{K}=1+\frac{5\pi }{8}\xi \left( s_{1}-s_{2}\right) ^{2}\left( \frac{h(e)}{\Delta \omega }\right) \, ,
\end{equation}
while for Kaluza-inspired theories, Eqs.(\ref{bd-tauxenertensbin})
and (\ref{ed-tauxenervectbin}), lead, to first order in the vector
charges, to\begin{equation}
\label{co-ratiov}
\left\langle \frac{\dot{T}^{VT}}{\dot{T}^{RG}}\right\rangle _{K}=\left\langle \frac{\dot{E}^{VT}}{\dot{E}^{RG}}\right\rangle _{K}=1+\frac{5\pi }{2}\left( \frac{v^{5}_{1}}{c}-\frac{v^{5}_{2}}{c}\right) ^{2}\left( \frac{h(e)}{\Delta \omega }\right) \, .
\end{equation}
The parameter \begin{equation}
\label{co-1}
\Delta \omega =\frac{6\pi GM}{ac^{2}(1-e^{2})}
\end{equation}
 stands for the general relativistic periastron precession in radians
per revolution, while \begin{equation}
\label{co-2}
h(e)\equiv \frac{1+e^{2}/2}{1+73/24e^{2}+37/96e^{4}}
\end{equation}
 is the reduction factor for eccentric orbits. The crucial parameter
is therefore \( h(e)/\Delta \omega  \): the more circular \( (h(e)\rightarrow 1) \)
and the less relativistic the orbit \( (\Delta \omega \rightarrow 0) \),
the more important the dipole radiations (relative to the quadrupole
ones). However, for maximum eccentricity \( (e=1) \), the reduction
factor \( h(e) \) is roughly \( 1/3 \) such that its impact on the
question of dipole dominance is not dramatic. On the other hand, too
small a relativistic precession would correspond to a vanishing and
therefore unobservable orbital decay. For further reference, the high
eccentricity \( (e\simeq 0.62) \) and relatively large periastron
angular shift (\( \Delta \omega \simeq 6.5\, 10^{-5} \) or \( \dot{\omega }\simeq 4.2\, ^{\circ }yr^{-1} \))
of \( PSR\, B1913+16 \) \cite{Tay93} are not considered as optimal
parameters for dipole dominance. 

Notice that these theoretical criteria selecting which binaries could
give the best constraints on dipole radiations are analogous to those
defined for testing explicitly the strong equivalence principle in
the strong field regime. It has been shown indeed \cite{Dam-Sch91}
that small-eccentricity long-orbital-period binary pulsars are also
the best laboratories in which a polarized motion (towards the center
of our galaxy) due to the strong equivalence principle violation can
be measured. However, in the light of the suggested alternative between
scalar and vector gravitational interactions, any constraint resulting
from such a direct analysis of the anomalous motion due to an equivalence
principle violation is in principle ambiguous. Indeed, the weakness
of the orbital polarization of a binary system (say, pulsar or Earth-Moon)
might be attributed to an accidental cancellation between the scalar
and vector perturbations rather than to the smallness of both the
scalar and vector charges. Such an ambiguity is absent from an indirect
analysis of dipole radiations since, in this case, the scalar and
vector emission rates simply add. Hence, any bound on dipole radiations
from binary pulsars provides a limit on \( (v^{5}_{1}/c-v^{5}_{2}/c)^{2} \)
and \( \xi (s_{1}-s_{2})^{2} \).

\subsection{Vector dipole in binary pulsars}

As already noticed, we don't know what really determines the vector
charge \( (v^{5}/c) \) of individual stars. However, its origin can
be understood as follows. The extra dimensional momentum of elementary
particles propagating in the compactified space is quantized. The
allowed values for this {}``transverse'' momentum actually define
the masses of the associated Kaluza-Klein excitations in the effective
\( 4 \)-dimensional theory \( (q_{(4+\delta )}^{2}=q_{4}^{2}-m_{n}^{2}c^{2}) \).
Hence, one might envisage the possibility that such excitations be
trapped into stars through gravitational accretion, just as WIMP's
ought to be. The vector charge for celestial bodies would then be
defined by the fraction of the total mass \( m \) due to these Kaluza-Klein
particles \( (v^{5}/c=\sum m_{KK}/m) \). But a dynamical model is
needed in order to estimate such a fraction in neutron stars.

\subsection{Scalar dipole in neutron star - white dwarf binaries}

The sensitivity \( (s) \) of each body being related to its gravitational
binding energy, the nature of the companion star is crucial for scalar
dipole radiations. In particular, the fact that neutron stars have
essentially identical sensitivities \( (s_{ns}\simeq 0.2) \) leaves
little hope for discriminating between general relativity and Brans-Dicke
with double neutron star binaries. Let us therefore turn to another
class of binaries.

Neutron star - white dwarf binaries are particularly interesting systems
when searching for dipole radiation effects since a white dwarf is
much less compact \( (s_{wd}\simeq 10^{-3}) \) than a neutron star.
Moreover, their orbital eccentricity is expected to be much smaller
if the observed neutron star is the first-born star and was recycled
by mass transfer from its companion \cite{Kas00}. So, the determination
of the orbital period decrease rate of a neutron star - white dwarf
binary pulsar might strongly constrain the scalar coupling of gravitation.
For illustration, we focus on the particular case of a recently discovered
system, \( PSR\, J1141-6545 \), which seems to be promising in terms
of relativistic orbital parameter measurements. Given \cite{Kas00}
its eccentricity \( (e\simeq 0.17) \) and periastron angular shift
(\( \Delta \omega \simeq 5.0\, 10^{-5} \) or \( \dot{\omega }\simeq 5.3\, ^{\circ }yr^{-1} \)),
Eq.(\ref{co-ratios}) implies

\begin{equation}
\label{co-3}
\left\langle \frac{\dot{T}^{ST}}{\dot{T}^{RG}}\right\rangle _{|(1141-6545)}\simeq 1+1.4\, 10^{3}\xi \, .
\end{equation}
An agreement to within \( 10\% \) with the general relativity prediction%
\footnote{The total mass of the system is \( M\simeq 2.300\, M_{\odot } \).
The mass function gives the bounds \( m_{p}\lesssim 1.348\, M_{\odot } \)
and \( m_{c}\gtrsim 0.968\, M_{\odot } \) for the mass of the pulsar
(which actually corresponds to the statistical value for neutron star
masses in known binary pulsars) and of the companion respectively.
Those figures lead to a prediction for the orbital period decrease
rate in general relativity of order \( \dot{T}^{RG}\simeq -3.881\, 10^{-13}\, s/s \). 
} would already give a constraint on the effective scalar coupling
\( \xi  \), of the same order of magnitude as the best current limits
given by solar system experiments. A more optimistic precision of
\( 1\% \) would provide us with the limit\begin{equation}
\label{co-4}
\xi \leq 10^{-5}\, \, \, \, \, (\omega \geq 10^{5})\, ,
\end{equation}
far better than the best current \( \xi \leq 1.5\, 10^{-4} \) constraint
\cite{Wil00} and reaching the precision level expected from future
satellite experiments dedicated to probing scalar gravitational couplings.
Needless to say that an indirect measurement of dipole radiations
would establish the existence of new gravitational charges.

\section{Conclusion}

Until now, the observed rates of relativistic orbital decay for binary
pulsars are consistent with a pure quadrupole emission of gravitational
waves, as predicted by general relativity. However, the appearance
of dipole radiations is quite generic in effective models inferred
from quantum theories that unify gravity with the other fundamental
interactions. In a scalar-tensor theory of gravitation, a natural
source for such radiations is the sensitivity \( (s) \) associated
with compact bodies. We have shown how more precise measurements of
the orbital parameters for peculiar binary pulsars consisting of a
neutron star and a white dwarf would strongly constrain the scalar
gravitational coupling \( (\xi ) \) of the Brans-Dicke model. Nevertheless,
if orbital decay rates ever suggest a non-vanishing anomalous coupling,
we will never be able to disentangle the minimal scalar-tensor extension
of Einstein's theory from a Kaluza-inspired vector-tensor one characterized
by an extra dimensional charge \( (v^{5}/c) \). The following interchange
between the hypothetical gravitational charges\[
\sqrt{\xi }\frac{s}{2}\, \leftrightarrow \, \frac{v^{5}}{c}\]
leads indeed to identical dipole rates. Only a direct detection of
gravitational waves could distinguish between longitudinal (scalar)
and transverse (vector) polarizations, and would consequently fix
the nature of the new charge involved.

\pagebreak

\textbf{Acknowledgements} \foreignlanguage{american}{}

We wish to thank Georgi Dvali, John Miller, Herman J.Mosquera Cuesta
and Lisa Randall for interesting discussions and helpful comments.
One of the authors (YW) is supported by the Fonds National de la Recherche
Scientifique (FNRS, Belgium).

\end{document}